\documentclass[letterpaper,english,prb,preprint,superscriptaddress]{revtex4}

\usepackage{graphics}

\makeatletter

\providecommand{\LyX}{L\kern-.1667em\lower.25em\hbox{Y}\kern-.125emX\@}

\makeatother
\begin{document}

\preprint{This line only printed with preprint option}

\title{NMR evidence for inhomogeneous density oscillations in the CuO chains of Ortho-II YBCO6.5}

\author{Z. Yamani}\email{Zahra.Yamani@nrc.gc.ca}
\altaffiliation{Permanent Address: NRC, CNBC, Chalk River, Canada} \affiliation{Department of Physics,
University of Toronto, Toronto, ON M5S 1A7, Canada }

\author{B.W. Statt}
\affiliation{Department of Physics, University of Toronto, Toronto, ON M5S 1A7, Canada }

\author{W.A. MacFarlane}\altaffiliation{Permanent Address:
Department of Chemistry, University of British Columbia, Canada} \affiliation{Department of Physics,
University of Toronto, Toronto, ON M5S 1A7, Canada }

\author{Ruixing Liang}
\affiliation{Department of Physics and Astronomy, University of British Columbia, Vancouver, BC V6T 1Z1,
Canada}
\author{D.A. Bonn}
\affiliation{Department of Physics and Astronomy, University of British Columbia, Vancouver, BC V6T 1Z1,
Canada}
\author{W.N. Hardy}
\affiliation{Department of Physics and Astronomy, University of British Columbia, Vancouver, BC V6T 1Z1,
Canada}

\date{\today}

\begin{abstract}

Nuclear magnetic resonance (NMR) measurements of CuO chains of  detwinned Ortho-II YBa\( _{2} \)Cu\( _{3}
\)O\( _{6.5}\) (YBCO6.5) single crystals reveal unusual and remarkable properties. The chain Cu resonance
broadens significantly, but gradually, on cooling from room temperature. The lineshape and its temperature
dependence are substantially different from that of a conventional spin/charge density wave (S/CDW) phase
transition. Instead, the line broadening is attributed to small amplitude static spin and charge density
oscillations with spatially varying amplitudes connected with the ends of the finite length chains. The
influence of this CuO chain phenomenon is also clearly manifested in the plane Cu NMR.

\end{abstract}
\maketitle

Common to all cuprate superconductors, the CuO$_2$ planes have been the focus of much scrutiny for their
essential role in the phenomenon of high temperature superconductivity (HTSC), including in the YBa\( _{2}
\)Cu\( _{3} \)O\(_{6+x}\) (YBCO6+x) family, whose layered structure also contains layers of
quasi-one-dimensional (quasi-1D) CuO chains. In addition to acting as a charge reservoir for the planes in
YBCO6+x, the chains have their own interesting properties due to their low dimensionality. In fact after
almost two decades, there is still no consensus on either the ground state or the low energy excitation
spectrum of the chains in YBCO6+x~\cite{ChainsGS-unknwon}. The behavior of quasi-1D correlated electrons has
been the subject of intense study over the past couple of decades from both theoretical and experimental
points of view, owing to the diversity of their electronic/magnetic properties as well as their amenability to
exact calculation~\cite{LowD}. For example, insulating S=\( 1\over 2 \) Heisenberg antiferromagnetic (AF)
copper-based systems (with chain, ladder and plane topology) have been investigated recently in relation to
the HTSC cuprates~\cite{DagRice}. Due to the difficulty in doping these compounds, the main results have been
to establish the detailed magnetic phase diagram, e.g. in Sr$_2$CuO$_3$~\cite{Takigawa97}. Doped AF quasi-1D
CuO chains in YBCO6+x are an interesting alternative to these commonly studied insulating chains. In HTSC
cuprates, the static AF insulating ground state of two dimensional CuO$_2$ planes is destroyed easily by
doping mobile carriers into these planes~\cite{zha96}. It is also important to investigate whether the
magnetic properties of CuO chains play an equally crucial role in determining the properties of adjacent doped
planes in YBCO6+x cuprate.

There are several oxygen ordered phases of YBCO6+x at certain fractional stoichiometries x. One such structure
is the Ortho-II phase of YBCO6.5 which has full CuO chains alternating with empty chains. To the extent that
the CuO chains are weakly coupled to the CuO$_2$ planes, the quasi-1D nature of the chains makes them
susceptible to the formation of S/CDW's. The effective dimensionality of the CuO chains in YBCO6+x is
determined by a hopping matrix element for mobile charges between the chains and planes (a process similar to
the interlayer coupling which has also been the focus of much recent debate in relation to HTSC~\cite{caxis}).
In low dimensional metallic systems, the formation of a S/CDW state is due to the topological enhancement of
Fermi surface nesting, which becomes perfect in the limit of 1D. Although the relevance of a Fermi liquid
picture is not at all clear in the cuprates, charge oscillations have indeed been observed in the CuO chains
by nuclear quadrupole resonance (NQR)~\cite{Grevin00} in YBCO7 and STM~\cite{Derro} in YBCO6+x. In YBCO7, the
Ortho-I phase of YBCO6+x, the chains and planes are apparently coupled strongly enough that the charge
oscillations in the chains induce charge oscillations in the planes of similar magnitude~\cite{Grevin00}. Thus
the present work on highly ordered stoichiometry Ortho-II YBCO6.5 provides an important opportunity to examine
not only a more isolated CuO chain than YBCO7, but also the chain-plane coupling effects in the underdoped
pseudogap regime, where the CuO$_2$ planes' response to perturbations is substantially
different~\cite{impurity}.

NMR spectroscopy is a unique probe in studying the microscopic properties of the CuO\( _{2} \) planes and CuO
chain layers both separately as well as with interactions, due to the ``site sensitivity'' of NMR, i.e. each
crystallographically inequivalent Cu site in the unit cell produces a separate resonance, whose lineshape and
spin relaxation properties reflect the local electromagnetic environment of that site. In this letter, we
present a detailed NMR study of the static properties of the full chain Cu sites in Ortho-II YBCO6.5. The
results demonstrate that the ground state of the chains is inhomogeneous in both electronic spin and charge
density but {\it cannot} be described by a conventional S/CDW ordered state. Additionally, we find that the
spin (charge) modulations in the chains induce small but measurable inhomogeneity in the planes.

High quality Ortho-II YBCO6.5 single crystals, T$_c$=62 K and $\Delta$T$_c$=0.6 K, were grown by a flux method
in BaZrO\( _{3} \) crucibles~\cite{UBCOrtho-II}. The crystals, in the form of ($ab$-plane) platelets several
millimeters on a side and a fraction of a millimeter thick ($c$-direction), were mechanically detwinned. X-ray
diffraction measurements indicate typical uninterrupted chain lengths of about 120
\textbf{b}~\cite{UBCOrtho-II}. Conventional spin-echo measurements were made using a homebuilt NMR
spectrometer with the field, {\bf H}$_0$, oriented along each of the crystallographic axes for various
temperatures and field strengths~\cite{Zahra02}. Because of the small size of the samples, considerable
averaging was essential to obtain reasonable signal-to-noise. The NMR spectrum observed for each Cu isotope
consists of four lines, corresponding to the four inequivalent Cu sites in the unit cell: full chain, empty
chain, planar Cu adjacent to either the full chain or empty chain, hereon denoted by Cu(1F), Cu(1E), Cu(2F)
and Cu(2E), respectively (see Fig. 1 in~\cite{Zahra02} for the crystal structure). Line assignments are based
on a comparison of the Knight shift and quadrupolar parameters obtained for these four sites to the known
values in YBCO6 and YBCO7~\cite{Zahra02}. To our knowledge, these are the narrowest and best-resolved
resonances observed for any oxygen depleted YBCO6+x material, an indication of the high quality of the
samples. At low temperatures, the central transition linewidths for the planar sites Cu(2) is of order 0.1 MHz
and that of the empty chain site Cu(1E) is somewhat less. In contrast, the Cu(1F) sites have a full width of
order of 1 MHz at low temperatures. The much broader resonance observed for Cu(1F) implies a remarkable
inhomogeneity in its local environment compared to the other sites. It is this inhomogeneity that is the main
subject of this study.

\begin{figure}
{\centering \resizebox*{.8\columnwidth}{!}{\includegraphics{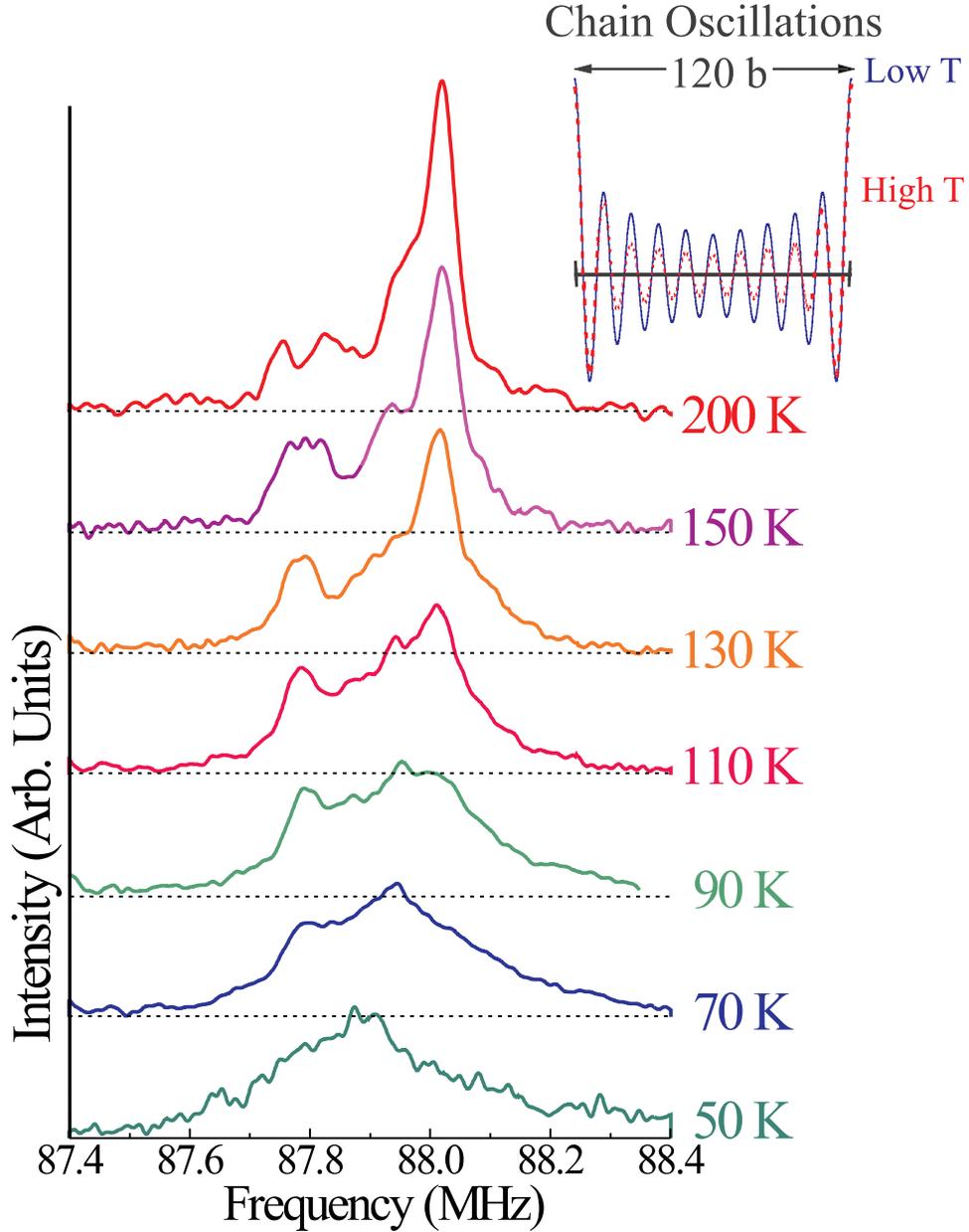}}
\par}
\caption{\label{Figure1} Temperature dependence of \protect\( ^{63}\protect \)Cu(1F) lineshape measured at 7.7
T with {\bf{H}}$_0 ||${\bf{a}}. Inhomogeneous density oscillations are depicted for low (solid line) and high
temperatures (dash line) in the inset. }
\end{figure}

The $^{63}$Cu(1F) lineshape for the applied field perpendicular to the chains ({\bf{H}}$_0 ||${\bf{a}}) is
shown as a function of temperature in Fig.~\ref{Figure1}(a). At low temperatures the line is broad. With
increasing temperature the broad line \textit{gradually} narrows while simultaneously a narrow line appears
and increases in relative intensity at the expense of the broad line. A similar temperature dependence has
also been observed for the applied field along the \textbf{b} and \textbf{c} axes. At the highest temperature
for which data has been collected (250 K) most of the intensity is in the narrow line, but there is still a
remnant of the broad line.

As CuO chains in Ortho-II YBCO6.5 are low dimensional, one might suggest the observed broadening at low
temperatures is due to a transition to a S/CDW state. For a simple homogeneous 1D incommensurate S/CDW state,
the NMR lineshape consists of two singularities separated by a frequency difference associated with the peaks
and troughs of the density waves~\cite{ButzCDW}. The singularities appear at the onset of S/CDW state and
their separation (a measure of the order parameter) increases on cooling until the ordered phase is fully
developed. Our data (see Fig.~\ref{Figure1} as an example), however, indicates that two singularities expected
from a simple homogeneous 1D incommensurate S/CDW state are absent in the Cu(1F) lineshape. In addition, the
smooth evolution of the lineshape with temperature (30 to 250 K) is inconsistent with a homogeneous density
wave phase transition characterized by a well-defined critical temperature T$_{DW}$~\cite{ButzCDW}. In
principle, one might expect that T$_{DW}$ could depend on chain length for any particular CuO segment;
however, the average segment length is so large, that most of them would already be in the homogeneous limit.
Hence, based on both the Cu(1F) lineshape and the temperature dependence of the measured spectra, we rule out
such a transition taking place in CuO chains in Ortho-II YBCO6.5.

We find, on the other hand, that the Cu(1F) lineshape \textit{is} consistent with presence of an spatially
varying amplitude oscillation induced by the chain ends depicted in the inset of Fig.~\ref{Figure1}. We
attribute the observed broad component of the Cu(1F) line to regions of the chains with large amplitude static
oscillations, while the narrow component at high temperatures is identified with regions where this amplitude
is small or zero, probably due to dynamic averaging. The absence of the narrow line at low temperatures
indicates that the oscillations have spread over the entire chain. The three-fold coordinated Cu site at the
chain end must have a different local charge than other chain Cu sites as it only has one neighbouring oxygen
in the chain instead of the usual two. Chain ends are thus effectively impurities in the chains. In a
conventional metal, due to the sharp cut-off in occupied conduction electron states at the Fermi surface, a
charged impurity leads to a spatially oscillating static screening cloud of conduction electrons, the
so-called Friedel oscillations~\cite{Berthier78}. Similarly, a magnetic defect in a metal gives rise to an
oscillating spin polarization of the screening conduction electrons, i.e. RKKY oscillations. There are
analogous effects in low dimensional {\it insulating} magnetic systems. Theoretical~\cite{Eggert95} and
experimental~\cite{Takigawa97} studies of insulating S=\( 1\over 2 \) 1D Heisenberg AF chains reveal
oscillations in the susceptibility near the chain ends. Thus one may also expect the doped S=\( 1\over 2 \)
chains in YBCO6.5 to have similar spin oscillations induced by the chain ends. Recently, Morr and
Balatsky~\cite{Morr01} have studied the electronic structure near impurities in chains of YBCO6+x coupled to
the planes and have indeed found spatially varying amplitude spin/charge oscillations in the electronic
density of states. Our experimental result confirms the presence of such oscillations in the CuO chains of
Ortho-II YBCO6+x.

\begin{figure}
{\centering \resizebox*{.8\columnwidth}{!}{\includegraphics{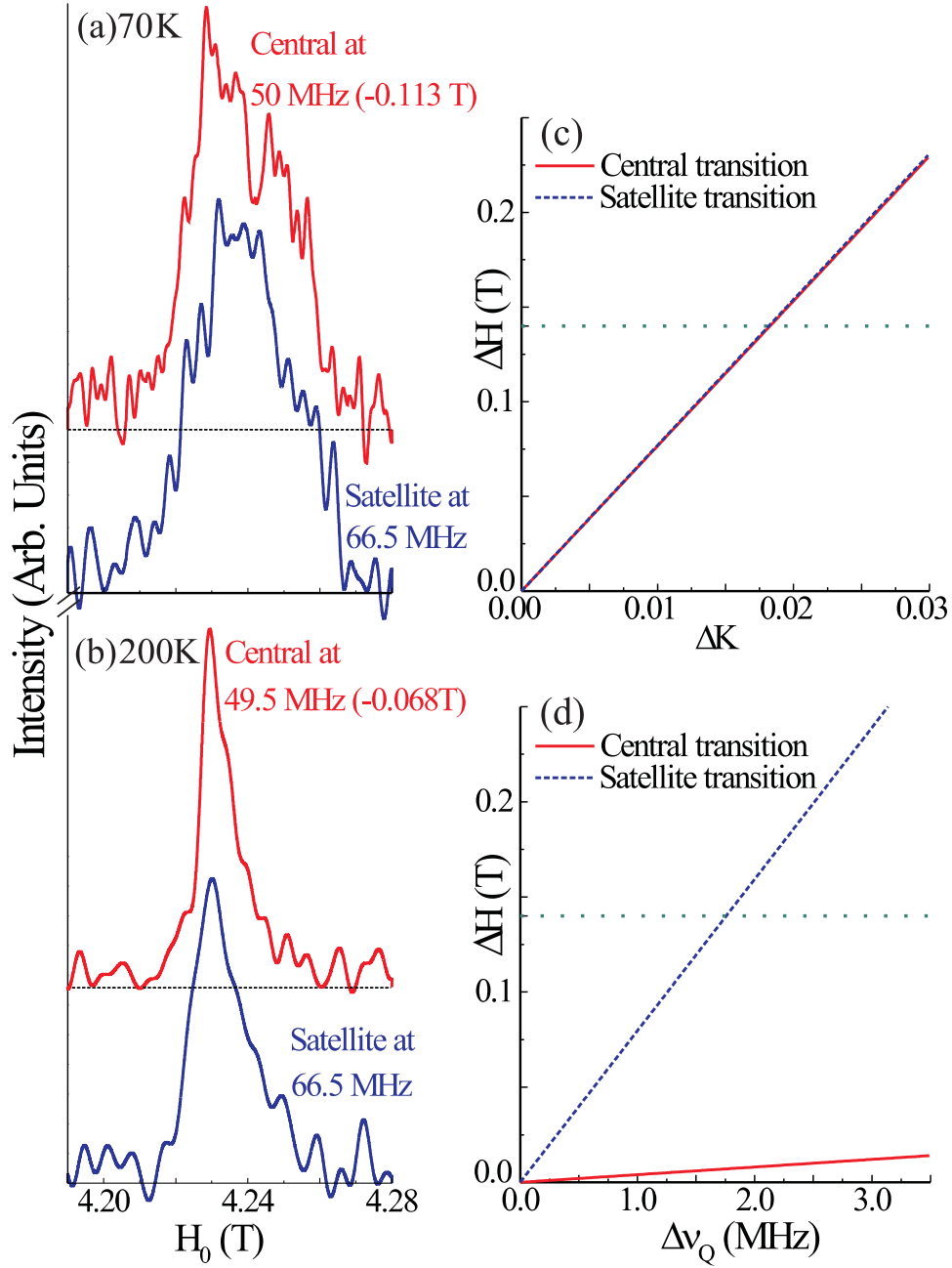}}
\par}
\caption{\label{Figure2} Central and satellite $^{63}$Cu(1F) spectra taken at 4.2 T at 70 K (a) and 200 K (b)
with {\bf{H}}$_0 ||${\bf{b}}. The central transition spectra are shifted (by the amount shown on the graphs)
to roughly align their center of gravity. The numerical simulations of the Cu(1F) resonance widths as a
function of Knight shift (c) and quadrupolar (d) distributions. The horizontal dashed-line in (c) and (d)
shows the maximum extent of the line observed for {\bf{H}}$_0 ||${\bf{c}}.}
\end{figure}

In general, the broadening of the Cu(1F) line can be due to inhomogeneity in {\it both} the local magnetic
field (through the local magnetic susceptibility, to the local magnetic shift) and the electric field gradient
(through the quadrupolar effects on the resonance). While the Knight shift distribution is the same first
order for both transitions, the quadrupolar broadening of the central transition is second order in nuclear
quadrupole frequency, $\nu_Q$, whereas it is first order for the satellite transitions~\cite{Zahra02}. Hence
by comparing the lineshapes of the central and satellite transitions, one can establish limits on the extent
of the distribution of the various quadrupolar and Knight shift parameters. Fig.~\ref{Figure2} shows typical
Cu(1F) spectra observed at low and high temperatures for both central and satellite transitions. It is clear
that the central and satellite transitions exhibit approximately the same the linewidths. Our numerical
simulations~\cite{Zahrathesis} further show that even large distributions in the quadrupolar parameters {\it
cannot} simultaneously reproduce the linewidths of both the central and satellite transitions (see
Fig.~\ref{Figure2} (c) and (d)). Thus the majority of the broadening of the Cu(1F) resonance comes from a
magnetic (Knight) shift distribution, with a much smaller contribution from quadrupolar broadening due to a
distribution in the charge density. The simulations indicate that a distribution of 0.1-0.25 MHz in \( \nu
_{Q} \) can explain the observed linewidths of satellite and central transitions for all field orientations.
Therefore, we do have evidence of {\it charge} oscillations, but our results show that they are very small in
amplitude. This is {\it not} because $\nu_Q$ is insensitive to the local charge, in fact systematic studies of
the role of the Cu ion's valence on $\nu_Q$ have found that $\nu_Q$ is extremely sensitive, varying by about
20 MHz per hole~\cite{Ken01}. Using this scaling, the distribution of \( \nu _{Q} \) here corresponds to a
static charge oscillation amplitude of about 0.5-1.25$\%$ of an electron per site.

We now consider the distribution in the Knight shift. There are two contributions to the Knight shift
$K$=$K_{s}$+$K_{orb}$ where \( K_{s} \) is the Knight shift due to the local spin susceptibility and \(
K_{orb} \) is the orbital shift (temperature independent shift due to a local electron screening of the
magnetic field). It is known from previous measurements~\cite{Takigawa91} of Cu(1F) in YBCO7 powder samples
that $K^{\bf {a}}_{orb}$$\simeq$1.2$\%$, $K^{\bf {b}}_{orb}$$\simeq$0.4$\%$ and $K_{s}$$\lesssim$0.2$\%$ at 70
K. In contrast, the maximum extent of the broadening of our central line is $\simeq$0.8$\%$ along \textbf{a}
and $\simeq$1.$3\%$ along \textbf{b} at 70 K, 7.7 T. Thus, since the orbital shift has the wrong symmetry, it
is doubtful that a distribution of $K_{orb}$ is the predominant source of broadening. \( K_{s} \) is too small
by an order of magnitude~\cite{Zahra02} to account for these linewidths. The hyperfine coupling constants, on
the other hand, do have the correct symmetry for all directions. Using estimates for the on-site and
transferred hyperfine coupling constants from Mila and Rice~\cite{MilaRice}, ($A$+2$B$)$_{\text {a}}$=80
KOe/$\mu_{B}$ and ($A$+2$B$)$_{\text {b}}$=120 KOe/$\mu_{B}$, we find that a 1.5$\%$ maximum polarization of
the local moments can account for our spectra, assuming the chain local moments have the same magnitude as
those in the planes, i.e. 0.6~\( \mu _{B} \)~\cite{Hyperfine}.

\begin{figure}[tbh]
{\centering \resizebox*{1\columnwidth}{!}{\includegraphics{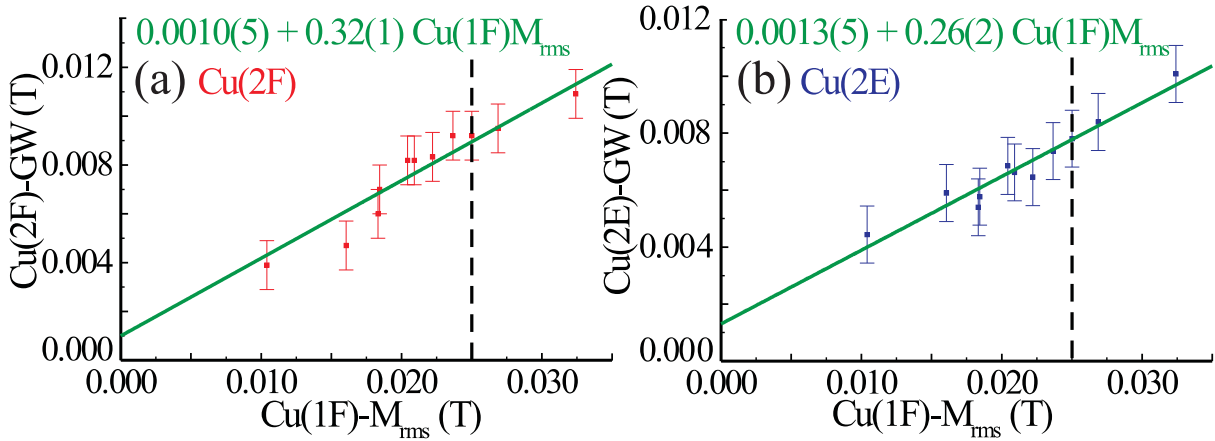}}
\par}
\caption{\label{Figure3} A linear relation is found between the Gaussian widths of the planar Cu(2F) and
Cu(2E) sites and Cu(1F) M$_{rms}$, (a) and (b) respectively, with {\bf{H}}$_0 ||${\bf{b}}. The dash line is
positioned at T$_c$. Similar linear behavior is also observed for the other field orientations.}
\end{figure}

Finally, we discuss the effects of chain oscillations on plane Cu sites. The broadening of the plane Cu
resonances follows that of Cu(1F), but is smaller in magnitude for Cu(2F) and smaller yet for Cu(2E). The
similarity of the temperature dependence of the plane Cu broadening to Cu(1F) broadening is evident in plots
of the plane Cu widths vs. the Cu(1F) rms moment~\cite{rmsM}, M$_{rms}$, with temperature as an implicit
parameter (see Fig.~\ref{Figure3} for {\bf{H}}$_0 ||${\bf{b}} for an example). Linear behavior found for all
field orientations, shows clearly that they all originate from the same phenomenon, namely the inhomogeneity
of the full CuO chains. Additionally the decreasing magnitude of this effect for sites progressively further
away from the CuO chains is strong evidence that the origin of the line broadening lies in the CuO chains.
Other independent sources of broadening are reflected in the small nonzero intercepts. This is in contrast to
YBCO7 where the $\nu_Q$ linewidth of both chains and plane sites suddenly increases below the superconducting
transition \textit{with the same magnitude}~\cite{Grevin00}. This qualitatively different behavior is likely a
consequence of a stronger chain-plane coupling in YBCO7, which is also reflected in its less anisotropic
electrical conductivity. In addition, unlike YBCO7, where the linewidths of both the chain and plane sites are
constant on cooling until below the superconducting T$_c$, for Ortho-II YBCO6.5 large but gradual increase in
the linewdith is observed on cooling. It has been suggested that in YBCO7 the Coulomb interaction between
charge carriers is responsible for suppressing the CDW order, until the CuO$_2$ planes become superconducting.
If this is the case, then the gradual increase indicates that a similar suppression effect of the charge/spin
oscillations is not evident in Ortho-II YBCO6.5.

We have investigated with NMR a system of weakly coupled chains and planes, namely Ortho-II YBCO6.5 and
present strong evidence that standing waves of spin (and to a lesser extent charge) density oscillations with
spatially varying amplitude form in the finite length chains at low temperatures. We find the oscillations are
reduced gradually with increasing temperature. This inhomogeneous electronic state in the chains clearly
influences the microscopic electronic structure in the planes. It is therefore important to understand this
phenomenon in order to properly distinguish between behaviour arising in the 1D chains from the effects
inherent to CuO$_2$ plane effects, such as the possible stripe phase.

We thank D.K. Morr, K. Samokhin, I. Vekhter and M.B. Walker for useful discussions. Financial support from
NSERC of Canada is gratefully acknowledged.


\begin{thebibliography}{10}
\bibitem{Hyperfine} This estimate is calculated based on the relation between the hyperfine field and the
effective moment: $A$+2$B$=$h_{AF}$/$\mu_{eff}$.
\bibitem{rmsM} Since the Cu(1F) lineshape appears to be complex, the rms moment, M$_{rms}$, is numerically calculated as a
representative of the linewidth.
\end{thebibliography}
\end{document}